\documentclass{aa}  
\usepackage{float}
\usepackage{natbib}
\usepackage{graphicx}
\usepackage{txfonts}
\usepackage{xspace}
\usepackage{xcolor}
\usepackage{lipsum}
\usepackage{subcaption}         
\usepackage{lscape}             
\usepackage{placeins}           
\usepackage[colorlinks=true,linkcolor=blue]{hyperref}%
             
\newcommand{\numax}{\ensuremath{\nu_{\rm max}}\xspace}
\newcommand{\dnu}{\ensuremath{\Delta\nu}\xspace}
\newcommand{\teff}{\ensuremath{T_{\rm eff}}\xspace}
\newcommand{\logg}{\ensuremath{\log g}\xspace}
\newcommand{\feh}{\ensuremath{[\mathrm{Fe/H}]}\xspace}

\newcommand{\msun}{\ensuremath{M_\odot}\xspace}
\newcommand{\rsun}{\ensuremath{R_\odot}\xspace}

\hypersetup{
    colorlinks=true,
    linkcolor=blue,
    citecolor=blue,
    urlcolor=blue
}
\begin{document}

   \title{CORALIE Radial Velocity Search for Companions around Evolved Stars (CASCADES)}

   \subtitle{VI: Revised masses and homogeneous estimation of the stellar parameters of 583 red giants}

%
%
%

   \author{E. Fontanet\inst{1}\corrauth{emile.fontanet@unige.ch}        
        \and S. Udry\inst{1}
        \and D. Ségransan\inst{1}
        \and P. Eggenberger\inst{1}
        \and P. Figueira\inst{1,2}
        \and J. P. Faria\inst{1}
        }

   \institute{Observatoire Astronomique de l’Université de Genève, Chemin Pegasi 51b, 1290 Versoix, Switzerland
   \and Instituto de Astrofísica de Andalucía-CSIC, Glorieta de la Astronomía s/n, 18008 Granada, Spain}

   \date{Received \today}

 
  \abstract
   {Giant-star radial-velocity surveys allow us to study how the exoplanet population varies with stellar mass and evolution. However, obtaining accurate estimates of stellar masses for such stars is challenging, as methods like isochrone fitting are prone to large errors due to the degeneracy associated with the closeness of isochrones.}
   {We provide a homogeneous re-derivation of the stellar parameters of the CASCADES giant-star survey. This includes parameters obtained from spectral energy distribution fitting, as well as an update of the stellar masses using asteroseismology.} 
   {We use TESS light curves to measure the frequency of maximum oscillation power, \numax, for all the stars when applicable. We re-extract the stellar radii of the survey by fitting broadband photometry with ARIADNE, a spectral energy distribution (SED) fitting package, using priors from spectroscopy for the effective temperature, metallicity and \logg. By using scaling relations, we compute revised estimates of the stellar masses of targets with a \numax measurement.}
   {We obtain new estimates of mass for 583 out of the 640 stars of the survey. These revised masses are typically 15 to 20\% lower (median difference of -0.35\, \msun) than the ones obtained by isochrone fitting, in particular for low-metallicity stars for which this effect is stronger. These new masses impact the overall mass distribution of the sample, whose median decreases from 2.11 \msun to 1.65 \msun.}
   {These revised masses provide the basis for future occurrence-rate analyses aiming to study the link between host-star mass and planet occurrence. They also highlight the uncertainties associated with the determination of masses for evolved stars.}


   \maketitle
   \nolinenumbers

\section{Introduction}
With the increasing number of exoplanet detections, our understanding of exoplanet demographics has improved substantially. Exoplanets are now found over a wide range of periods, masses, and host stars. A question that remains of actuality is how the occurrence rate of exoplanets depends on
host-star mass and metallicity \citep{johnsonGiantPlanetOccurrence2010, reffertPreciseRadialVelocities2015}. To answer this, one needs to find exoplanets around more massive hosts, for which planet occurrence remains less well constrained. Studying more massive stars, in the 1.5 - 5 \msun range, is however a challenging task for stars on the main sequence \citep{borgnietExtrasolarPlanetsBrown2019}. Since they are hotter and rotate faster, achieving the radial-velocity (RV) precision to detect exoplanets orbiting them is particularly hard. Red giants, however, have a slower rotation rate and cooler surface, making them more suitable for RV surveys. Studying intermediate-mass giant stars, whose progenitors may have been A to F stars on the main sequence, is thus a proxy to study this population of stars. In this context, several surveys have been launched to look for exoplanets around these evolved hosts \citep{frinkDiscoverySubstellarCompanion2002,satoPlanetaryCompanionsThree2008,  jonesStudyImpactPostMS2011, wittenmyerPanPacificPlanetSearch2011b}.

While giant stars facilitate obtaining precise RV measurements compared to their main-sequence progenitors, they also bring some difficulties with them. One of the main difficulties with these stars is the precise determination of their masses. This is a crucial problem, as the interpretation of planet occurrence around evolved stars relies critically on host-star masses. The main issue is that stellar masses obtained from spectroscopy and isochrone fitting face strong degeneracy for stars spanning this area of the Hertzsprung-Russell (HR) diagram. Both more massive first-ascent red giant branch (RGB) stars and less massive, core-helium-burning red clump stars pass through the same diagram loci, making it difficult to distinguish them without additional constraints. This issue is known, and was already addressed in some previous works \citep{lloydRETIREDPLANETHOSTS2011,schlaufmanEvidenceTidalDestruction2013,ghezziRetiredStarsRevisited2018,mallaBenchmarkingSpectroscopicMasses2024}.

To counter this issue, asteroseismology can help as an additional method to add constraints on the stellar parameters. Solar-like oscillations can be measured in red giants' photometric time series. Two global seismic parameters, the frequency of maximum oscillation power (\numax) and the large frequency separation (\dnu), which is the average frequency spacing between consecutive radial oscillation modes, can be used to estimate the stellar masses and radii. While this has been done for several Kepler or CoRoT samples \citep{sreenivasSimpleMethodMeasure2024a,miglioProbingPopulationsRed2009}, the increasing number of sectors of the Transiting Exoplanet Survey Satellite (TESS) mission \citep{rickerTransitingExoplanetSurvey2015} now allows this technique to be applied to many bright red giants \citep{stelloTESSAsteroseismologyKepler2022}, including those outside of the continuous viewing zone. While obtaining a robust value of \dnu remains challenging for stars with a limited number of sectors, measuring the position of \numax, combined with an independent estimate of stellar radius and effective temperature, allows us to estimate the stellar mass through the scaling relation $M \propto \numax \cdot R^2 \cdot T_{\rm eff}^{1/2}$ \citep{brownDetectionPossiblePMode1991,kjeldsenAmplitudesStellarOscillations1995,chaplinAsteroseismologySolarTypeRedGiant2013} (Sect.\, \ref{sec:seismic_scaling_rel}, Eq.\, \ref{eq:seismic_mass_scaling}). This can be done homogeneously over a full sample of red giants, providing revised estimates of stellar masses that are less affected by the degeneracies linked to isochrone fitting.

In this work, we revise the stellar parameters of the CORALIE Radial-Velocity Search for Companions Around Evolved Stars (CASCADES), a long-term RV program targeting 640 G- and K-type red giants. We derive updated spectral energy distribution (SED) parameters using ARIADNE \citep{vinesARIADNEMeasuringAccurate2022}, extract \numax values from TESS light curves and compute revised stellar masses using seismic scaling relations for a subsample of 583 stars. 

The paper is organized as follows. Section \ref{sec:sample_stellar_params} presents the CASCADES sample and stellar parameters obtained from SED fitting. Section \ref{sec:astero_analysis} describes the TESS photometric data and its processing, as well as the extraction of \numax values. Section \ref{sec:stellar_masses} presents the revised masses and how they compare with previous estimates, and we discuss the results and their implications for the survey in section \ref{sec:discuss}.

\section{Sample and stellar parameters}
\label{sec:sample_stellar_params}
\subsection{The CASCADES sample}
\begin{figure}[t]
        \centering
        \includegraphics[width=\columnwidth]{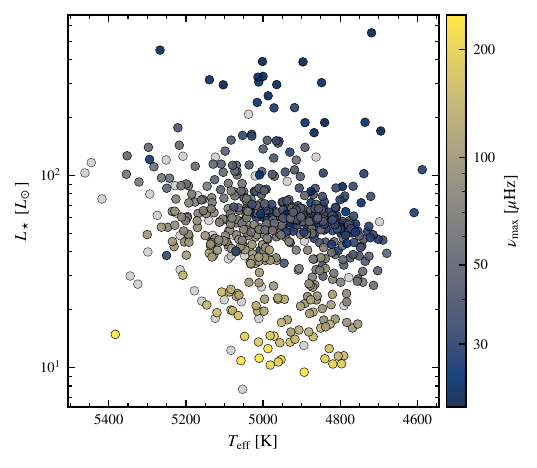}
        \caption{HR diagram of the stars of the CASCADES survey, color-coded by the value of \numax measured from TESS light curves. Stars for which we did not manage to measure \numax are colored in light gray.}
        \label{fig:hr_diag}%
    \end{figure}
The CASCADES sample consists of 640 G- and K-type red giant stars that have been followed in RV for about 20 years with the CORALIE spectrograph mounted on the Leonhard Euler Swiss telescope in La Silla, Chile. The survey was first presented in \citet{ottoniCORALIERadialvelocitySearch2022}, along with a list of stellar parameters for a subsample of 620 stars, which included stellar radii, masses, metallicities, effective temperatures, and surface gravities, among others.

While spectroscopic effective temperatures and metallicities are generally well constrained, stellar masses of red giants remain difficult to determine from spectroscopy and isochrone fitting alone. The reason for this is that these stars occupy a crowded region on the HR diagram, where stars spanning a wide range of masses are packed together. This creates a degeneracy between more massive stars ascending the red giant branch and less massive core-helium-burning red-clump stars, which already underwent the helium flash. These degeneracies can then lead to overestimated stellar masses from isochrone fitting \citep{schlaufmanEvidenceTidalDestruction2013,mallaAsteroseismicMassesFour2020,buldgenCORALIERadialVelocity2022,mallaBenchmarkingSpectroscopicMasses2024}.

Out of the 640 stars of the original sample, we removed 16 stars that have been identified as double-lined spectroscopic binaries. For these systems, the composite spectra and unresolved photometry can bias both the SED-derived parameters and the interpretation of the TESS power spectra. These targets are also not considered in the exoplanet occurrence analysis. On top of this, we were unable to extract a reliable value of \numax for 36 other stars. These cases were rejected after visual inspection of the power spectra. For these stars, the smoothed power spectra did not show a well-defined broad power excess, the uncertainties on the \numax estimates were too large, or the power spectra were dominated by coherent harmonic structures inconsistent with the expected solar-like oscillation pattern. This leaves a total of 588 targets for which we provide a value of \numax. For five of these stars, the available broadband photometry did not allow us to obtain reliable estimates of effective temperature and stellar radius from SED fitting (Sect.~\ref{sec:ariadne}), leaving 583 stars for which we provide revised masses.

\begin{figure*}[t]
        \centering
        \includegraphics[width=\textwidth{}]{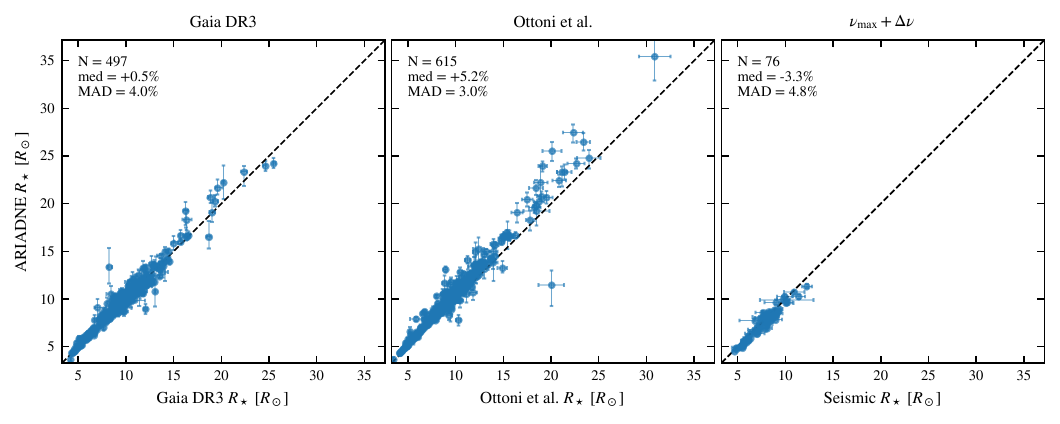}
        \caption{Comparison of the values of stellar radii from ARIADNE with Gaia DR3 radii \citep{creeveyGaiaDataRelease2023} (left), with the values of \citet{ottoniCORALIERadialvelocitySearch2022} (middle) and with the seismic radii for stars with a \dnu measurement (right). The diagonal dashed line corresponds to 1:1 agreement.}
        \label{fig:radius_comp}%
    \end{figure*}
    
\subsection{SED parameters from ARIADNE}
\label{sec:ariadne}
In addition to the revision of stellar masses using asteroseismology, we decided to also update stellar parameters such as effective temperature (\teff), stellar radii, stellar luminosities, and other standard SED quantities using ARIADNE \citep{vinesARIADNEMeasuringAccurate2022}, an SED-fitting package. ARIADNE\footnote{\url{https://github.com/jvines/astroARIADNE}} is a Python code that performs Bayesian model averaging (BMA) of different stellar atmosphere models using broadband photometry data from several surveys. We ran ARIADNE on the full sample homogeneously using the PHOENIX \citep{husserNewExtensiveLibrary2013},
BT-family \citep{allardModelsVerylowmassStars2012},
Kurucz \citep{kuruczATLAS9StellarAtmosphere1993}, and CK04
\citep{castelliNewGridsATLAS92003} model grids for a total of 100 000 samples per target. We used ARIADNE's default priors for all quantities except for \teff, \logg  and \feh, for which we used a Gaussian prior centered on the spectroscopic estimates of \citet{ottoniCORALIERadialvelocitySearch2022}, with standard deviations equal to the reported uncertainties. These estimates were derived using high-resolution spectra obtained with the CORALIE spectrograph. The default priors include a uniform prior on the stellar radius between 0.05\,R$_\odot$ and 100\,R$_\odot$, and a Gaussian prior for the distance, centered on the distance estimate of \citet{bailer-jonesEstimatingDistanceParallaxes2018} with a standard deviation equal to five times the larger of the two quoted distance uncertainties. For the extinction A$_V$, the default is a uniform prior between zero and the maximum line-of-sight extinction from the dust map of \citet{1998ApJ...500..525S}. A full description of the algorithm is available in \citet{vinesARIADNEMeasuringAccurate2022}. We systematically removed all the outliers in the photometric data using ARIADNE's built-in outlier identifying routine. We could not obtain revised SED parameters for five stars of the sample, for which sufficient photometry was not available.

The quantities which matter the most in the context of this paper are the stellar radii and effective temperatures, which are used in Eq.\, \ref{eq:seismic_mass_scaling} in Sect.\, \ref{sec:seismic_scaling_rel} to estimate the stellar mass from \numax. A comparison of the values of stellar radii obtained with ARIADNE with those from Gaia Data Release 3 (DR3) and from \citet{ottoniCORALIERadialvelocitySearch2022} is shown in Fig.\, \ref{fig:radius_comp}. The median fractional difference between ARIADNE and Gaia DR3 radii \citep{creeveyGaiaDataRelease2023}, defined as the median of $(R_{\rm ARIADNE} - R_{\rm Gaia})/R_{\rm Gaia}$  is +0.5\%, with a median absolute deviation (MAD) of 4\%. The comparison with the radii from \citet{ottoniCORALIERadialvelocitySearch2022} shows a median offset of +5.2\%, with a MAD of 3\%, with ARIADNE yielding slightly larger radii on average. Overall, this supports the adopted radius estimates, while showing a small systematic difference with respect to the previous analysis by \citet{ottoniCORALIERadialvelocitySearch2022}.

We also compared the ARIADNE effective temperatures with the spectroscopic values reported by \citet{ottoniCORALIERadialvelocitySearch2022}. The median fractional difference is $-0.09\%$, with a MAD of $0.19\%$. This close agreement is expected given the use of the spectroscopic parameters as Gaussian priors in the SED analysis. The SED-derived parameters therefore represent a refinement incorporating the existing spectroscopic constraints, rather than a fully independent determination.

For a subset of 76 stars with sufficiently clear oscillation spectra, we also derived radii from \dnu and \numax measurements obtained with pySYD \citep{chontosPySYDAutomatedMeasurements2022}, along with the \teff from spectroscopy, by combining Eq.\, \ref{eq:numax_logg_scaling} and Eq.\, \ref{eq:dnu_scaling}. These radii are not used in the adopted mass estimates, but provide an additional consistency check on the values of radii obtained with ARIADNE. While we were only able to extract \dnu for a small sample of stars, mostly at lower radii, the estimated radii agree with those of ARIADNE, with a median fractional difference of -3.3\% and MAD of 4.8\% (taking the seismic radius as reference).

As can be seen in Eq.\, \ref{eq:seismic_mass_scaling}, our estimate of the stellar mass depends on $(R/R_\odot)^2$. It is therefore essential to have a reliable estimate of the stellar radius. The posterior uncertainties on $R$ and \teff\ were propagated when deriving the final stellar masses.

\section{Asteroseismic analysis}
\label{sec:astero_analysis}
\subsection{TESS light curves and data preparation}
We downloaded publicly available light curves from the TESS mission \citep{rickerTransitingExoplanetSurvey2015}
from the Mikulski Archive for Space Telescopes (MAST) using \texttt{astroquery.mast}\footnote{\url{https://astroquery.readthedocs.io/en/stable/mast/mast.html}} \citep{ginsburgAstroqueryAstronomicalWebquerying2019}. When available, we used the Science Processing Operations Center (SPOC) pre-search data conditioning light curves \citep{jenkinsTESSScienceProcessing2016}. If no SPOC light curve was available, we used those from the Quick-Look Pipeline (QLP) \citep{huangPhotometry10Million2020}. When both 120-second and 20-second SPOC light curves were available, we used the 120-second cadence data, since the expected \numax values of our red-giant targets are well below the Nyquist frequency of the 120-second data.

We then applied a basic detrending to the data. First, we performed a sigma-clipping of outliers at 5 sigma. We then estimated the slowly varying baseline  using a Savitzky-Golay filter
\citep{savitzkySmoothingDifferentiationData1964} with a smoothing window of 10 days, and divided the original light curve by this baseline before computing the power spectrum. Only points with a QUALITY flag of 0 were kept. We only used corrected fluxes, either \texttt{PDCSAP\_FLUX} or \texttt{KSPSAP\_FLUX}. Sectors for which only \texttt{SAP\_FLUX} was available were excluded, since these uncorrected light curves were found to introduce unreliable \numax measurements in our analysis.

\begin{figure}[t]
        \centering
        \includegraphics[width=\columnwidth]{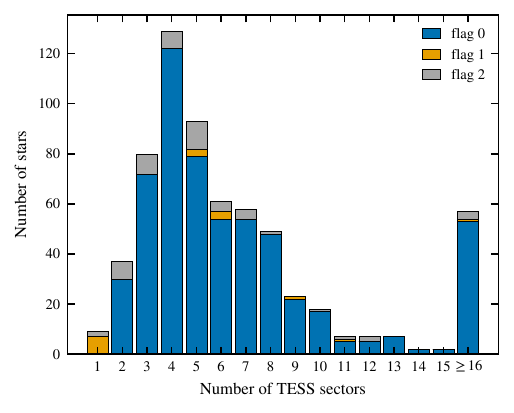}
        \caption{TESS coverage of the stars of the CASCADES survey. The histogram displays the number of sectors available per star that we used in our analysis. Each TESS sector spans approximately 27 days, and sectors for a given star may be non-consecutive. The colors represent the flags assigned to the extraction of the \numax value with 0 being accepted, 1 being accepted with reserve and 2 being rejected.}
        \label{fig:tess_coverage}%
    \end{figure}

This resulted in a total of 5007 sectors for the 640 stars of the survey (respectively 4615 for the 583 stars for which we provide an update on the mass). The distribution of the number of sectors per star can be seen in Fig.\, \ref{fig:tess_coverage}. The median number of sectors per star is 5, and only 9 stars have a single sector of data. As TESS sectors span approximately 27 days, the median coverage is about 135 days per star. For a given star, the sectors may be non-consecutive.

\subsection{Extraction of \numax}

\begin{figure*}[t]
        \centering
        \includegraphics[width=\textwidth]{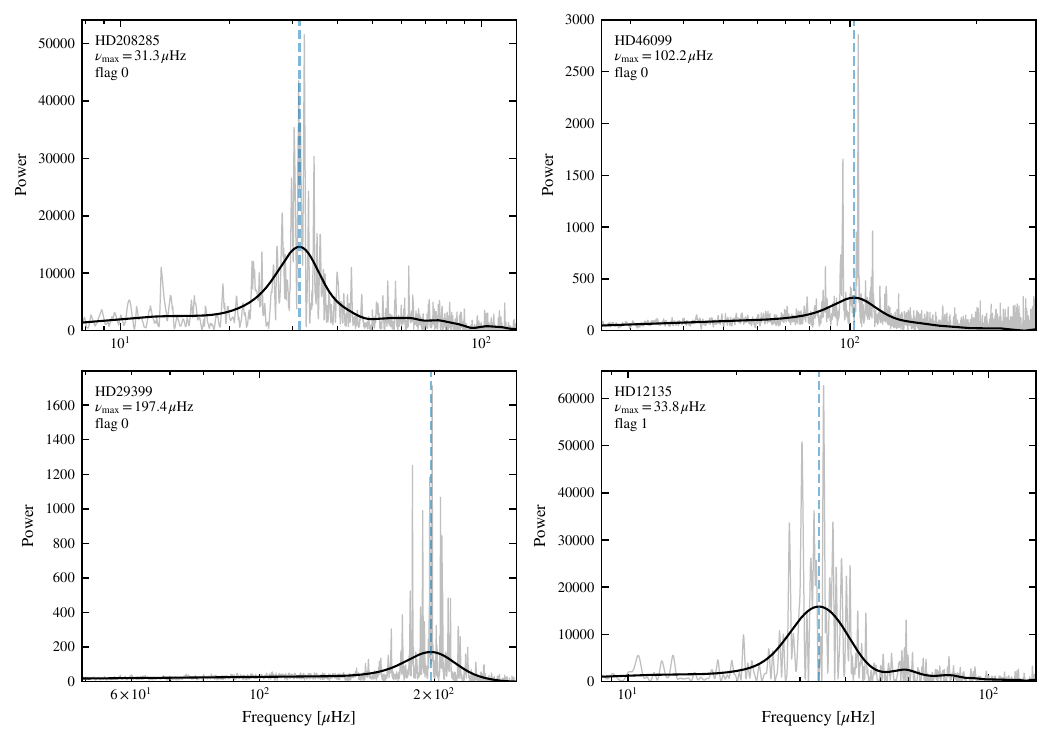}
        \caption{Example of power spectra and the corresponding \numax values for three stars with low, medium and high values of \numax obtained with the nuSYD Python package. The bottom-right panel shows an example of a flag-1 star, corresponding to a lower-confidence measurement of \numax. For this star, the flag was set to "1" because only one sector of data was available for the extraction. The gray spectrum is the median spectrum obtained by combining the individual power spectra of all available sectors, and the smooth black line shows the smoothed spectrum. The dashed vertical line represents the uncorrected \numax position.}
        \label{fig:numax_ex}%
    \end{figure*}
The values of \numax were extracted from the light curves using the nuSYD Python package \citep{sreenivasSimpleMethodMeasure2024a}. nuSYD allows to measure values of \numax by creating a smoothed, background-corrected power spectrum. Whenever several sectors were available, we computed the power spectra of each sector on a common frequency grid using nuSYD, computed the median spectrum, and measured the value of \numax on this master spectrum. We adopted the median rather than the mean of the individual power spectra to reduce the influence of sector-specific artifacts. Combining the power spectra in this way rather than at the time-series level also does not require phase coherence between different sectors. We also computed and stored all the individual values of \numax per sector, as well as the ones obtained by stitching consecutive sectors when there were any. In the stitching case, we first detrended each sector individually, and then concatenated the light curves in chronological order, preserving the original timestamps. For each star, we provided an initial guess for \numax by inverting Eq.\,\ref{eq:seismic_mass_scaling} using the masses, radii, and effective temperatures from \citet{ottoniCORALIERadialvelocitySearch2022}. This value was used only as an initial guess for the extraction and not as a prior on the final \numax measurement. As reported by \citet{sreenivasSimpleMethodMeasure2024a}, the raw \numax estimate returned by nuSYD is affected by a small systematic bias introduced by the smoothing and background-correction procedure. We therefore adopted the bias-corrected \numax values calibrated by \citet{sreenivasSimpleMethodMeasure2024a}.

Targets with ambiguous power excesses, strong disagreement between values of \numax per sector or poor-quality light curves were flagged and/or excluded from the clean \numax-based mass sample. We used the following flags: 0 for good \numax extraction, 1 for stars with moderate sector-to-sector disagreement and 2 for stars with no clear \numax or excessive sector-to-sector disagreement. We quantified the sector-to-sector scatter using the MAD of the individual sector \numax measurements divided by their median. Stars with a fractional scatter between 20\% and 40\% were assigned flag~1, while those with a scatter above 40\% were assigned flag~2. Single-sector stars were automatically assigned a flag of 1, because their sector-to-sector consistency could not be assessed. Stars without any identified concerns were assigned flag~0. All the spectra were then visually assessed after the first round of flagging. A few examples of \numax extractions can be seen in Fig.\, \ref{fig:numax_ex}. 

\subsection{Validation and comparison with literature}
In our final sample, we removed all the stars with a flag of 2, and kept those with flags 0 and 1. In the table containing updated parameters, we added a column indicating the flag, as well as a column indicating the reason for the flag (for the case of flag 1 stars). \numax values with a flag of 1 are still considered to be acceptable, but can be less reliable and are typically lower signal to noise. An example of a star yielding a \numax with quality flag of 1 can be seen in Fig.\, \ref{fig:numax_ex} (in this case a single-sector measurement).

To assess the quality of our approach, we compared the values of \numax that we obtained with those from other studies with shared targets. We found 400 common stars with \citet{honCloseinGiantPlanet2023}, 253 with \citet{zhouDetectionSolarlikeOscillations2024a}, 28 with \citet{sreenivasGlobalAsteroseismology192026} and 22 with \citet{mallaBenchmarkingSpectroscopicMasses2024}. As some stars are present in more than one catalog, this results in a total of 465 unique stars. 

\begin{figure*}[t]
        \centering
        \includegraphics[width=\textwidth]{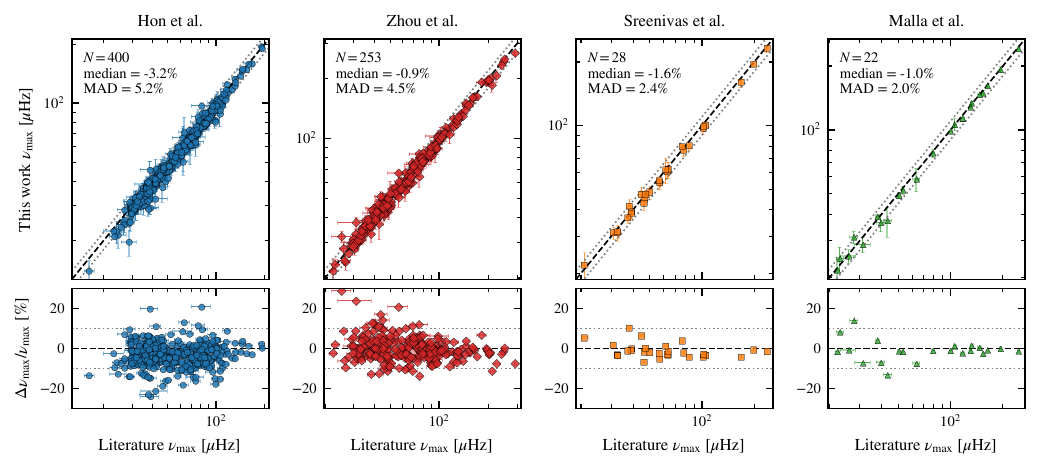}
        \caption{Values of \numax obtained in this work compared to the literature results of  \citet{honCloseinGiantPlanet2023}, \citet{zhouDetectionSolarlikeOscillations2024a}, \citet{sreenivasGlobalAsteroseismology192026} and \citet{mallaBenchmarkingSpectroscopicMasses2024}, from left to right.}
        \label{fig:numax_comp}%
    \end{figure*}

Fig.\, \ref{fig:numax_comp} shows the comparison between the results of our analysis and those of the aforementioned works. The median fractional differences between our \numax measurements and the literature values are -3.2\%, -0.9\%, -1.6\%, and -1.0\% for the comparisons with \citet{honCloseinGiantPlanet2023}, \citet{zhouDetectionSolarlikeOscillations2024a}, \citet{sreenivasGlobalAsteroseismology192026}, and \citet{mallaBenchmarkingSpectroscopicMasses2024}, respectively. The corresponding scatter values, measured using the MAD, are 5.2\%, 4.5\%, 2.4\%, and 2\%. This agreement supports the reliability of the \numax estimates adopted in this work.

\section{Revised stellar masses}
\label{sec:stellar_masses}
\subsection{Seismic scaling relations}
\label{sec:seismic_scaling_rel}
Stellar masses were estimated from the asteroseismic scaling relation \citep{brownDetectionPossiblePMode1991,kjeldsenAmplitudesStellarOscillations1995,chaplinAsteroseismologySolarTypeRedGiant2013} combining the measured \numax values with the ARIADNE radii and effective temperatures:

\begin{equation}
\frac{\nu_{\rm max}}{\nu_{\rm max,\odot}}
\simeq
\frac{g}{g_\odot}
\left(\frac{T_{\rm eff}}{T_{{\rm eff},\odot}}\right)^{-1/2}
=
\frac{M}{M_\odot}
\left(\frac{R}{R_\odot}\right)^{-2}
\left(\frac{T_{\rm eff}}{T_{{\rm eff},\odot}}\right)^{-1/2}
\label{eq:numax_logg_scaling}
\end{equation}

\begin{equation}
\frac{M}{M_\odot}
=
\left(\frac{\nu_{\max}}{\nu_{\max,\odot}}\right)
\left(\frac{R}{R_\odot}\right)^2
\left(\frac{T_{\rm eff}}{T_{{\rm eff},\odot}}\right)^{1/2}
\label{eq:seismic_mass_scaling}
\end{equation}

\begin{equation}
\frac{\Delta\nu}{\Delta\nu_\odot}
\simeq
\left(\frac{M}{M_\odot}\right)^{1/2}
\left(\frac{R}{R_\odot}\right)^{-3/2}
\label{eq:dnu_scaling}
\end{equation}
With solar reference values of $\nu_{\max,\odot} = 3090\,\mu{\rm Hz}$, $\Delta\nu_\odot = 135.1\,\mu{\rm Hz}$, and $T_{{\rm eff},\odot} = 5772\,{\rm K}$.

We did not attempt to measure the large frequency separation $\Delta\nu$ for the full sample because robust $\Delta\nu$ extraction requires additional validation, especially for targets with short TESS baselines, complex spectral windows, or low signal-to-noise oscillation spectra. Instead, we use $\nu_{\max}$ as a surface-gravity constraint and combine it with independent SED radii. This choice trades some precision for homogeneity and robustness across the full survey sample. We thus used Eq.\, \ref{eq:seismic_mass_scaling} to compute masses for the 583 stars with both a \numax measurement of sufficient quality from our sample and a radius estimate from ARIADNE.

To estimate the uncertainty on the mass estimate, we built a mass posterior by evaluating Eq.\, \ref{eq:seismic_mass_scaling} on the posterior samples of effective temperatures and radii obtained with ARIADNE. Since nuSYD does not provide posterior samples for \numax, we drew \numax samples from a normal
distribution centered on the adopted \numax value and with a standard deviation given by our adopted \numax uncertainty. We found the formal nuSYD uncertainties to be small compared to the sector-to-sector scatter. We therefore adopted the sector scatter, defined as the Gaussian-scaled MAD between sectors, as the \numax uncertainty when three or more sectors were available. For stars with two sectors, we used half of the absolute difference between the two \numax values. For single-sector targets, we adopted the maximum between the formal nuSYD uncertainty and 5.2\% of the \numax value, corresponding to the largest MAD observed in the literature comparison shown in Fig.\,\ref{fig:numax_comp}. The reported mass and uncertainties correspond to the median and the 16th and 84th percentiles of the resulting mass posterior.

\subsection{Comparison with isochrone masses}

\begin{figure}[t]

        \centering
        \includegraphics[width=\columnwidth]{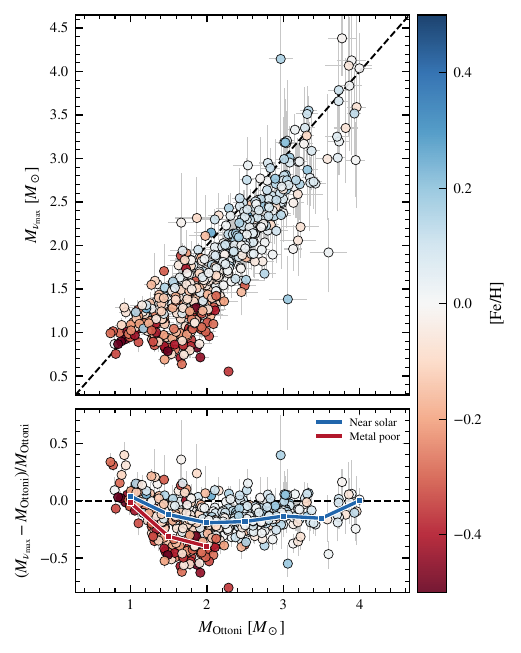}
        \caption{Top panel: Comparison of stellar masses obtained with isochrones (\citet{ottoniCORALIERadialvelocitySearch2022}) and from \numax measurements (this work). The circles are color-coded by stellar metallicity. Bottom panel: The fractional difference as a function of the isochrone mass. The blue and red curves show the median fractional differences in mass bins of 0.5\,\msun for near-solar ($-0.1\leq[\rm Fe/H]\leq0.1$) and metal-poor ($[\rm Fe/H]\leq -0.2$) stars, respectively. Only bins containing at least 5 stars are shown. The lower-panel error bars include only the \numax-based mass uncertainty.}
        \label{fig:mass_comp}%
    \end{figure}

A comparison of the revised masses of the 583 stars with the ones obtained by \citet{ottoniCORALIERadialvelocitySearch2022} using SPInS \citep{lebretonSPInSPipelineMassive2020} can be seen in Fig.\, \ref{fig:mass_comp}. While the two distributions show a coherent trend, a clear, non negligible offset can also be seen. Isochrone fitting seems to systematically provide larger estimates of the stellar masses compared to \numax-based masses, especially for low-metallicity targets. This discrepancy has already been reported in previous studies (e.g. \citet{mallaBenchmarkingSpectroscopicMasses2024}, \citet{schlaufmanEvidenceTidalDestruction2013}, \citet{buldgenCORALIERadialVelocity2022}).

Two groups can be identified from this comparison. First, the main sample consists of stars excluding the most discrepant low-metallicity targets. These stars show a more or less continuous offset between the two approaches, typically of 15-20\%, with many targets remaining compatible within the uncertainties. This includes the most massive stars in the sample (M$_{\rm{iso}}$ > 2 \msun) and, more generally, the more metal-rich stars.

The second group forms a cluster of low-metallicity stars at the lower-mass end of the distribution. These targets show a very significant difference in mass between the two approaches ($|\Delta$M/M$_\text{iso}|$ > 0.4, in many cases corresponding to differences larger than 0.7 \msun). The isochrone mass estimates of these stars are mainly in the 1.5 - 2 \msun range, while the \numax-based mass estimates place them in the 0.75 - 1.3 \msun mass range. Most of these stars are consistent with being low-mass red-clump/core-helium-burning stars (luminosity between 50 and 60 L$_\odot$ and \teff between 4700 K and 5000 K on the HR diagram from Fig.\, \ref{fig:hr_diag}), a possibility that was already mentioned by \citet{ottoniCORALIERadialvelocitySearch2022}. This behavior is consistent with the known degeneracy between low-mass core-helium-burning stars and more massive first-ascent RGB or secondary-clump stars in the HR diagram \citep{girardiSecondaryClumpRed1999}. While metallicity was taken into account in the mass estimates of \citet{ottoniCORALIERadialvelocitySearch2022}, the concentration of the largest discrepancies among metal-poor targets suggests that, in our sample, this degeneracy affected the metal-poor part of the sample more severely. The \numax measurement provides an additional constraint on $M/R^2$, and therefore on the stellar surface gravity, which helps to break this mass degeneracy. Most of these targets have a value of \numax between 20 and $\sim$ 40 $\mu$Hz, and a radius between 10 and 11 \rsun, as can be seen in Fig.\, \ref{fig:numax_rad_dist}. They can also be seen in Fig.\, \ref{fig:hr_diag} in the 50-60 L$_\odot$ range, color-coded by their value of \numax.

\section{Discussion}
\label{sec:discuss}
\subsection{Validity and limitations}
While the revised SED parameters obtained with ARIADNE (\teff, radii, luminosities, among others) are in very good agreement with the previous estimates by \citet{ottoniCORALIERadialvelocitySearch2022}, the update on the stellar masses that we propose substantially changes the overall mass distribution of the targets. This naturally raises the question of the validity of the new masses. 

One limitation of our work is that we only consider \numax and do not attempt to measure \dnu for the full sample. With a reliable measurement of \dnu, both the stellar mass and radius could be obtained directly from seismic scaling relations by using Eq.\, \ref{eq:dnu_scaling} and Eq.\, \ref{eq:seismic_mass_scaling}, together with an estimate of \teff. Our approach instead uses \numax as a surface-gravity constraint and combines it with the SED radii from ARIADNE. The extraction of \numax without \dnu was done without the goal of doing an in-depth analysis of individual targets, but rather to obtain a revised, homogeneous set of stellar parameters for the full sample. It is likely that a more thorough analysis could improve the estimate of the masses, but such an analysis is beyond the scope of this survey-level study. Moreover, obtaining robust \dnu measurements from TESS data is significantly more challenging than measuring \numax, especially for stars with a small number of sectors, complex spectral windows, or low signal-to-noise. Although we do not use \dnu for the full-sample mass determination, the subset of 76 stars for which we measured it provides an additional consistency check that the ARIADNE radii are consistent with seismic radii.

Despite this limitation, \numax provides a valuable seismic constraint because
it scales approximately with $g/\sqrt{T_{\rm eff}}$ (Eq.\, \ref{eq:numax_logg_scaling}) and therefore directly
constrains $M/R^2$ \citep{morelAssessingAccuracySurface2012}. The use of \numax combined with independently determined SED radii therefore provides an additional seismic constraint that is not available from spectroscopy and isochrone fitting alone.

The revised mass distribution can also be considered in the context of other red-giant populations. Our sample shows a concentration near $\numax\sim30\,\mu{\rm Hz}$ and $R\sim10\,\rsun$ (Fig.\,\ref{fig:numax_rad_dist}). \citet{miglioProbingPopulationsRed2009} showed, using CoRoT red giants and population-synthesis models, that the dominant population with $20 < \numax < 50\,\mu{\rm Hz}$ is associated with red-clump stars, with radii around 10 - 12\,\rsun and masses mainly between about $0.8$ and 1.8\,\msun. These similarities are consistent with a red-clump population in CASCADES, although we emphasize that the exact mass distribution of CASCADES depends on the survey selection function and cannot be directly inferred from the CoRoT population. A similar population was also observed in the Kepler red-giant sample by \citet{hekkerCharacterizationRedGiant2011}.

\begin{figure}[t]
    \centering
    \includegraphics[width=\linewidth]{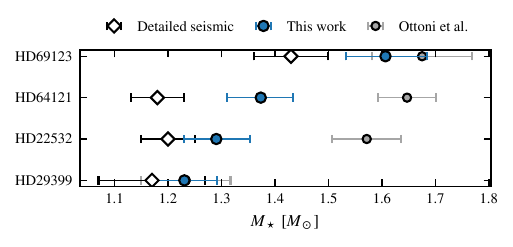}
    \caption{Comparison of the stellar masses from this work to the ones obtained from precise asteroseismic analysis by \citet{buldgenCORALIERadialVelocity2022} and \citet{pezzottiCORALIERadialvelocitySearch2022}. }
    \label{fig:precise_mass_comparison}
\end{figure}

As an additional validation, detailed asteroseismic masses of four stars of the survey were obtained in the past by \citet{buldgenCORALIERadialVelocity2022} and \citet{pezzottiCORALIERadialvelocitySearch2022}. Their estimates were obtained from more detailed analyses of the oscillation spectra of individual stars, rather than the survey-level approach that we adopt in this work. Figure\,\ref{fig:precise_mass_comparison} compares these detailed seismic estimates with our results and the isochrone masses from \citet{ottoniCORALIERadialvelocitySearch2022}. Our masses are closer to the detailed seismic estimates in all four cases. They agree within uncertainties for HD\,22532 and HD\,29399, but remain higher for HD\,64121 and HD\,69123. This further supports the overall revision of the stellar masses of the survey, while also illustrating the limitations of the survey-level approach for individual targets.

\subsection{Implications for the CASCADES planet survey}
\begin{figure}[t]
        \centering
        \includegraphics[width=\columnwidth]{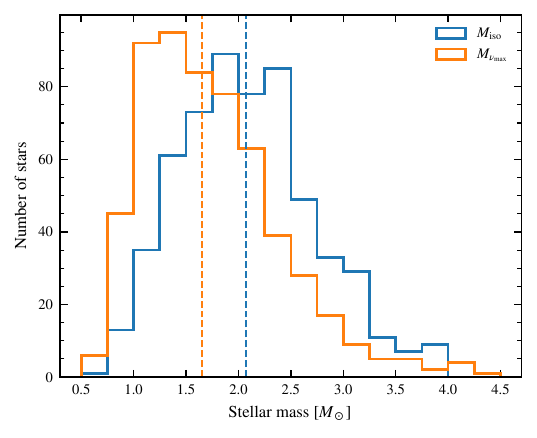}

        \caption{Mass distribution of the stars in the CASCADES sample before (blue) and after (orange) the mass update for the same 583 stars. The vertical lines indicate the median of each distribution.}
        \label{fig:mass_distrib}%
    \end{figure}

The revised masses obtained in the context of this work have important implications for the survey as a whole. The original distribution of stellar masses from \citet{ottoniCORALIERadialvelocitySearch2022} can be seen and compared to our update in Fig.\, \ref{fig:mass_distrib}. We observe that the peak of the distribution has clearly shifted toward lower masses, with a new median stellar mass of 1.65 \msun (versus 2.11 \msun before). 

Since CASCADES was started with the aim of studying the population of exoplanets around intermediate-mass stars, this update has a direct impact on the science case of the survey. The sample of stars with a mass larger than 1.5 \msun has gone from 505 to 342, substantially reducing the number of targets classified as intermediate-mass stars. Nevertheless, the revised sample still contains several hundred stars above 1.5\,\msun, and therefore remains suitable for occurrence-rate studies as a function of stellar mass.

Naturally, a change in host star masses also has an impact on the parameters of detected planets themselves. As of September 2026, there have been 12 detections in the context of the CASCADES survey \citep{ottoniCORALIERadialvelocitySearch2022,pezzottiCORALIERadialvelocitySearch2022,fontanetCORALIERadialvelocitySearch2025,figueiraCORALIERadialvelocitySearch2026a}. The change in the estimate of the host star mass thus yields a change in the estimate of the planet minimum mass. For RV-detected planets, assuming $M_{\rm p} \ll M_\star$, the inferred minimum mass scales with the host-star mass as $M_{\rm p}\sin i \propto M_\star^{2/3}$. For the planets analyzed by \citet{fontanetCORALIERadialvelocitySearch2025} and \citet{figueiraCORALIERadialvelocitySearch2026a}, adopting our revised stellar masses decreases the published minimum masses by approximately 0.3-16\%. The decreases remain below 10\% for all systems except HD\,102888. The planetary nature of the previously published companions is therefore not challenged by the revised stellar masses. For the four hosts with previously published masses from detailed asteroseismic analyses, we retain the published stellar and companion masses.

\section{Conclusions}
With the present work, we provide the following improvements to the CASCADES survey:
\begin{enumerate}
    \item a homogeneous revision of stellar parameters coming from SED with ARIADNE
    \item the extraction of \numax values for 588 stars using data coming from the TESS mission
    \item an update of the stellar masses using the \numax scaling relation based on the values of \numax and the updated radii and effective temperatures
\end{enumerate}
With this update, the median stellar mass of the sample decreases from 2.11\,\msun to 1.65\,\msun. The number of stars above 1.5\,\msun decreases from 505 to 342. The largest downward revisions occur preferentially for low-metallicity stars located in the red-clump region of the HR diagram, consistent with the known degeneracy affecting isochrone masses of evolved stars. The comparison with literature \numax catalogs shows good agreement, supporting the consistency of our \numax measurements with the literature. 

These revised masses provide a more robust basis for future CASCADES occurrence
rate studies as a function of host-star mass. They will also affect the inferred
minimum masses of RV-detected companions, although only weakly, since
$M_{\rm p}\sin i \propto M_\star^{2/3}$. More importantly, this work highlights the risk associated with relying solely on isochrone-based stellar masses for evolved stars.

\section{Data availability}
All the new stellar parameters, including revised masses, \numax values, TESS sectors used and flags will be made available in electronic format at the Centre de Données astronomiques de Strasbourg (CDS) upon publication.
\begin{acknowledgements}
This work has been carried out within the framework of the National Center of Competence in Research PlanetS supported by the Swiss National Science Foundation under grants $51NF40\_182901$ and $51NF40\_205606$. The authors also acknowledge financial support from the SNSF through the MINT project under grant $200020\_205010$. This paper includes data collected with the TESS mission, obtained from the MAST data archive at the Space Telescope Science Institute (STScI). Funding for US Institutions for the TESS mission is provided by the NASA Explorer Program. STScI is operated by the Association of Universities for Research in Astronomy, Inc., under NASA contract NAS5–26555. This work has made use of data from the European Space Agency (ESA) mission Gaia (https://www.cosmos.esa.int/gaia), processed by the Gaia Data Processing and Analysis Consortium (DPAC, https://www.cosmos.esa.int/web/gaia/dpac/consortium). Funding for the DPAC has been provided by national institutions, in particular the institutions participating in the Gaia Multilateral Agreement. PF acknowledges financial support from the Severo Ochoa grant CEX2021-001131-S funded by MCIN/AEI/10.13039/501100011033. PF is also funded by the European Union (ERC, THIRSTEE, 101164189). Views and opinions expressed are however those of the author(s) only and do not necessarily reflect those of the European Union or the European Research Council. Neither the European Union nor the granting authority can be held responsible for them. PE acknowledges support from the SNSF grant No 219745 (Asteroseismology of transport processes for the evolution of stars and planets).
\end{acknowledgements}

\bibliographystyle{bibtex/aa}
\bibliography{bibliography}

\begin{appendix}
\section{\numax and radius distribution}

\begin{figure}[h]
        \centering
        \includegraphics[width=\columnwidth]{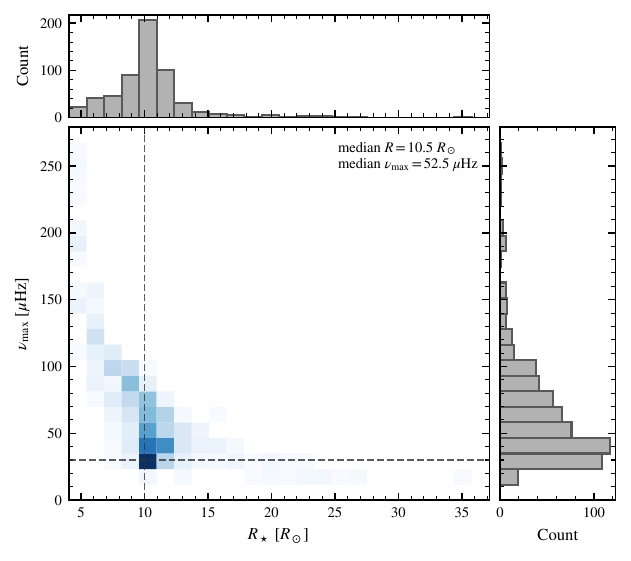}
        \caption{Joint distribution of stellar radii and values of \numax for the 583 stars with both a \numax measurement and a radius estimate. The horizontal and vertical lines indicate \numax = 30 $\mu$Hz and R = 10 \rsun, respectively, near the peak of the joint distribution.}
        \label{fig:numax_rad_dist}%
    \end{figure}

\end{appendix}
\end{document}